\documentclass[aps,pra,preprint,showpacs]{revtex4}
\usepackage{times}
\usepackage{nicefrac}
\usepackage{amsmath}
\usepackage{amsfonts}
\usepackage{amssymb}
\usepackage{amsthm}
\newcommand{\EV}[1]{\left\langle\Psi\left|#1\right|\Psi\right\rangle}

\begin{document}
\preprint{Version 2.0}

\title{Ground state wave function and energy of the lithium atom}

\author{Mariusz Puchalski}
\email[]{mpuchals@fuw.edu.pl}

\author{Krzysztof Pachucki}
\email[]{krp@fuw.edu.pl}

\affiliation{Institute of Theoretical Physics, Warsaw University,
             Ho\.{z}a 69, 00-681 Warsaw, Poland}


\begin{abstract}
Highly accurate nonrelativistic ground-state wave function and energy of the lithium atom
is obtained in the Hylleraas basis set. The leading relativistic corrections,
as represented by Breit-Pauli Hamiltonian, are obtained in fair agreement with 
the former results. The calculational method is based on the analytical
evaluation of Hylleraas integrals with the help of recursion relations. 
\end{abstract}

\pacs{31.25.Eb, 31.30.Jv 31.15.Pf, 02.70.-c}
\maketitle
\section{Introduction}

Theoretical predictions for the energy levels of light few-electron atoms
are much less accurate than for the hydrogenic systems. It is for two
reasons. The nonrelativistic wave function has to include 
electron correlations to a high degree of accuracy. This can be achieved
by using a Hylleraas basis set, but it is quite difficult 
to evaluate integrals with Hylleraas functions for three and more
electrons.  The second reason is the difficulty in the accurate treatment
of relativistic and radiative corrections. The commonly used 
Dirac-Coulomb Hamiltonian for few-electron atoms does not include
relativistic corrections properly as it cannot be derived from
quantum electrodynamic theory and its continuous spectrum  ranges from $-\infty$ to $+\infty$.
One of the possible approaches is the derivation of an effective Hamiltonian \cite{fw}
within the so called NRQED theory. Matrix elements 
of this Hamiltonian give exact correction to the energy at specified order in the
fine structure constant $\alpha$. However, this Hamiltonian becomes quite complicated
at higher orders and for example $m\,\alpha^6$ corrections has been obtained
for few low lying states of helium only \cite{singlet,triplet}, not for lithium nor
beryllium atoms.

Theoretical predictions for light hydrogen-like atoms are at present limited
by uncertainty in higher-order two-loop electron self-energy corrections \cite{yer}, 
which is a few kHz for the 1S state. For helium-like atoms predictions
are approximately $10^3$ times less accurate. Since, the nonrelativistic wave
function was computed very accurately using Hylleraas \cite{drake_he} or exponential basis
sets \cite{kor_he}, the uncertainty in its levels comes mainly from the unknown $m\,\alpha^7$
terms. These corrections are currently under investigation in the context
of helium $2^3P_J$ fine splitting.
For lithium atoms, the Hylleraas functions give very accurate
nonrelativistic wave function and energies \cite{yan_lit2},
but the precise  calculation of three-electron integrals
with Hylleraas functions is very time consuming \cite{king_lit, yan_lit1}, 
and so far no result for $m\,\alpha^6$ corrections have been obtained. For the beryllium atom
the most accurate results have been obtained with explicitly correlated
Gaussian functions \cite{komasa_be}. Although it was possible to calculate accurately 
the leading relativistic and QED corrections \cite{beqed}, 
the final accuracy is limited by the nonrelativistic energy.
Moreover, this basis cannot be used for higher order corrections since 
Gaussian wave functions do not fulfill the cusp condition.

So far the most accurate results for various states of the lithium atom were 
obtained by Yan and Drake in Ref. \cite{yan_lit2}. Here, we present 
even more accurate results for the lithium ground state,
as a demonstration of an analytic method
to compute the integrals with Hylleraas functions \cite{recursions}. This new method 
is based on recursion relations between integrals with different powers of
electron-nucleus and inter-electron distances,
which are fast and numerically stable for generating large basis sets.
Our result for the ground state energy 
\begin{equation}
E = -7.478\,060\,323\,904\,1(^{+10}_{-50})\,,
\label{02}
\end{equation}
is significantly below  the previous one, obtained in \cite{yan_lit2}, which is
$E=-7.478\,060\,323\,650\,3(71)$. As a further application
of the analytic approach, we obtain the leading relativistic
corrections to the binding energy by the calculation of the expectation value 
of Breit-Pauli Hamiltonian in Eq. (\ref{11}). For this we used recursion relations
for extended Hylleraas integrals with $1/r_{ij}^2$ and $1/r_i^2$ terms.
They have been derived in \cite{rec_sing} and in this work respectively.

In the next Section we construct the nonrelativistic wave function, similarly
to Ref. \cite{yan_lit2} and obtain the ground state nonrelativistic energy and
the wave function. In Sec. III
we compute the leading relativistic correction as given by the Breit-Pauli
Hamiltonian. In Sec. IV we derive recursion relations for 
Hylleraas integrals containing 
$1/r_i^2$ which among others, are necessary for relativistic matrix elements. 
In Sec. V we summarize our result
and present prospects for calculation of higher order terms as well as
the calculation of Hylleraas integrals for 4 and more electrons.

\section{Nonrelativistic wave function and energy}
In the construction of the wave function we closely follow the works of Yan and Drake
in \cite{yan_lit2}. The ground state wave function $\Psi$ is expressed as a linear combination
of $\psi$, the antisymmetrized product of $\phi$ and the spin function $\chi$
\begin{eqnarray}
\psi &=& {\cal A}[\phi(\vec r_1,\vec r_2,\vec r_3)\,\chi]\,,
\label{03}\\
\phi(\vec r_1, \vec r_2, \vec r_3) &=& e^{-w_1\,r_1-w_2\,r_2-w_3\,r_3}\,
r_{23}^{n_1}\,r_{31}^{n_2}\,r_{12}^{n_3}\,r_{1}^{n_4}\,r_{2}^{n_5}\,r_{3}^{n_6}\,,
\label{01} \\
\chi &=& \alpha(1)\,\beta(2)\,\alpha(3)-\beta(1)\,\alpha(2)\,\alpha(3)\,,
\label{04}
\end{eqnarray}
with all $n_i$ nonnegative integers and $w_i\in {R}_+$.
The matrix element of the Hamiltonian $H$
\begin{equation}
H = \sum_{a=1}^3\,\left(\frac{\vec p_a^{\,2}}{2}-\frac{Z\,\alpha}{r_a}\right)
+ \sum_{a>b=1}^3\,\frac{\alpha}{r_{ab}}\,,
\label{05}
\end{equation}
or of any spin independent operator can be expressed after eliminating spin variables, as
\begin{eqnarray}
\langle\psi|H|\psi'\rangle &=& \langle
2\,\phi(1,2,3)+
2\,\phi(2,1,3)-
\phi(3,1,2)-
\phi(2,3,1)-
\phi(1,3,2)-
\phi(3,2,1)|\nonumber \\ &&
 H\,|\phi'(1,2,3)\rangle\,.
\label{06}
\end{eqnarray}
In this way the calculation of this matrix elements is brought to 
Hylleraas integrals, namely the integrals with respect to $r_i$
of the form
\begin{eqnarray}
f(n_1,n_2,n_3,n_4,n_5,n_6) &=& \int \frac{d^3 r_1}{4\,\pi}\,
                               \int \frac{d^3 r_2}{4\,\pi}\,
                               \int \frac{d^3 r_3}{4\,\pi}\,
                               e^{-w_1\,r_1-w_2\,r_2-w_3\,r_3}
\nonumber \\ &&
r_{23}^{n_1-1}\,r_{31}^{n_2-1}\,r_{12}^{n_3-1}\,r_{1}^{n_4-1}\,r_{2}^{n_5-1}\,r_{3}^{n_6-1}\,,
\label{07}
\end{eqnarray}
with nonnegative integers $n_i$. These 
are performed analytically for $n_1, n_2,n_3=0,1$ \cite{remiddi} and by recursion
relations for larger $n_i$ using formulas derived in \cite{recursions}.

The total wave function is generated from all $\phi$ in Eq. (\ref{01})
with $n_i$ satisfying condition
\begin{equation}
\sum_{i=1}^6 n_i \leq \Omega\,,
\label{08}
\end{equation}
for $\Omega$ between 3 and 12. 
For each $\Omega$ we minimize energy with respect to the free parameters $w_i$  
in Eq. (\ref{01}). We noticed that the use of only one set of $w_i$'s
does not lead to accurate results, therefore, following Yan and Drake \cite{yan_lit2},
we divide the whole basis set into 5 sectors, each one with its own set of
$w_i$'s. This division goes as follows \cite{yan_lit2}
\begin{center}
\begin{tabular*}{4in}[t]{rlrr}
sector 1: & all $n_3$,      & $n_1=0$,      & $n_2=0$;     \\
sector 2: & all $n_3$,      & $n_1=0$,      & $n_2\neq 0$; \\
sector 3: & all $n_3$,      & $n_1\neq 0$,  & $n_2= 0$;    \\
sector 4: & $n_3 = 0$,      & $n_1\neq 0$,  & $n_2\neq 0$; \\
sector 5: & $n_3 \neq 0$,   & $n_1\neq 0$,  & $n_2\neq 0$; 
\end{tabular*}
\end{center}
To avoid numerical instabilities, within each sector we drop the terms
with $n_4>n_5$ (or $n_4<n_5$) and for $n_4=n_5$ drop terms with $n_1>n_2$ (or $n_1<n_2$).  
This division allows for a significant improvements of nonrelativistic
energies by optimization of all five sets of $w_i$'s.
These nonlinear parameters are obtained by Newton method of searching zeros
using analytic derivatives
\begin{equation}
\frac{\partial E}{\partial w} = 2\left\langle
\Psi\biggl|H\biggr|\frac{\partial\Psi}{\partial w}\right\rangle
- 2\,E\,\left\langle\Psi\biggl|\frac{\partial \Psi}{\partial w}\right\rangle\,.
\label{09}
\end{equation}

In the numerical calculations, we use sextuple precision for recursion
relations and quadruple precision for all other arithmetics
to obtain the wave function and the energy up to $\Omega=12$.
The results obtained for ground state energies are presented in Table I.
The penultimate row is a result of extrapolation to infinite length of the basis
set, and the last raw are previous results of Yan and Drake \cite{yan_lit2}.
\begin{table}[!hbt]
\caption{Ground state nonrelativistic energies and expectation values
         of Dirac $\delta$-functions obtained using Drachman formulae
         \cite{Dra81} for various basis length.}
\label{table1}
\begin{ruledtabular}
\begin{tabular}{rrlll}
$\Omega$ &No. of terms & $E(\Omega)$   & $\sum_a \delta^3(r_a)$        &
     $\sum_{a>b}\delta^3(r_{ab})$   \\ \hline
     $3$ &50   & -7.477\,981\,524\,089\,7 & 13.843\,446\,803\,98 & 0.544\,164\,351\,92  \\
     $4$ &120  & -7.478\,052\,334\,642\,2 & 13.842\,288\,641\,67 & 0.544\,331\,564\,16  \\
     $5$ &256  & -7.478\,059\,463\,915\,8 & 13.842\,509\,174\,63 & 0.544\,327\,870\,45  \\
     $6$ &512  & -7.478\,060\,208\,663\,7 & 13.842\,637\,966\,67 & 0.544\,325\,260\,63  \\
     $7$ &918  & -7.478\,060\,310\,362\,9 & 13.842\,606\,662\,38 & 0.544\,324\,788\,85  \\
     $8$ &1589 & -7.478\,060\,320\,507\,6 & 13.842\,608\,240\,76 & 0.544\,324\,697\,02  \\
     $9$ &2625 & -7.478\,060\,323\,450\,1 & 13.842\,610\,098\,57 & 0.544\,324\,629\,45  \\
     $10$ &4172& -7.478\,060\,323\,775\,0 & 13.842\,610\,698\,67 & 0.544\,324\,627\,57  \\
     $11$ &6412& -7.478\,060\,323\,861\,0 & 13.842\,610\,779\,19 & 0.544\,324\,631\,50  \\
     $12$ &9576& -7.478\,060\,323\,889\,7 & 13.842\,610\,781\,06 & 0.544\,324\,632\,05  \\
$\infty$ &$\infty$& -7.478\,060\,323\,904\,1$(^{+10}_{-50})$ & 13.842\,610\,783\,46(100) & 0.544\,324\,633\,96(50)  \\
  Refs. \cite{yan_lit2,yan_delta}&$\infty$& -7.478\,060\,323\,650\,3(71) &
     13.842\,609\,642\,(55) &  0.544\,329\,79(31)  \\
\end{tabular}
\end{ruledtabular}
\end{table}
The result for the nonrelativistic energy is significantly below the previous
estimate \cite{yan_lit2} and indicates that extrapolation to infinite basis length
does not always give the right result. 
In the same Table we present results for the Dirac $\delta$ functions,
which also differs from previous results in \cite{yan_delta}.
We observe, the the number of significant digits for Dirac $\delta$ is 
increased by using Drachman formulae \cite{Dra81}, namely
\begin{eqnarray}\label{Eijd}
4\pi\EV{\delta^3(r_{ab})}&=&
  2\EV{\frac{1}{r_{ab}}(E_\Psi-V)} -\sum_c\left\langle{
\vec \nabla}_c\Psi\left|\frac{1}{r_{ab}}\right|{\vec \nabla}_c\Psi\right\rangle,
\\
4\pi\EV{\delta^3(r_{a})}&=&
  4\EV{\frac{1}{r_{a}}(E_\Psi-V)} -2\,\sum_c\left\langle{
\vec \nabla}_c\Psi\left|\frac{1}{r_{a}}\right|{\vec \nabla}_c\Psi\right\rangle\,.
\end{eqnarray}
where $V$ is a total potential energy in Eq. (\ref{05}).

\section{Leading relativistic correction to binding energy}
The leading relativistic corrections to energy levels are given by the
expectation values of the Breit-Pauli Hamiltonian $H^{(4)}$. 
\begin{eqnarray}
H^{(4)} &=&\sum_a \biggl\{-\frac{\vec p^{\,4}_a}{8\,m^3} +
\frac{ \pi\,Z\,\alpha}{2\,m^2}\,\delta^3(r_a)
+\frac{Z\,\alpha}{4\,m^2}\,
\vec\sigma_a\cdot\frac{\vec r_a}{r_a^3}\times \vec p_a\biggr\}
\nonumber \\
&& +\sum_{a>b}\sum_b \biggl\{
-\frac{ \pi\,\alpha}{m^2}\, \delta^3(r_{ab})
-\frac{\alpha}{2\,m^2}\, p_a^i\,
\biggl(\frac{\delta^{ij}}{r_{ab}}+\frac{r^i_{ab}\,r^j_{ab}}{r^3_{ab}}
\biggr)\, p_b^j \nonumber \\ && -
\frac{2\, \pi\,\alpha}{3\,m^2}\,\vec\sigma_a
\cdot\vec\sigma_b\,\delta^3(r_{ab})
+\frac{\alpha}{4\,m^2}\frac{\sigma_a^i\,\sigma_b^j}
{r_{ab}^3}\,
\biggl(\delta^{ij}-3\,\frac{r_{ab}^i\,r_{ab}^j}{r_{ab}^2}\biggr)
+\frac{\alpha}{4\,m^2\,r_{ab}^3} \nonumber \\ &\times& \biggl[
2\,\bigl(\vec\sigma_a\cdot\vec r_{ab}\times\vec p_b -
\vec\sigma_b\cdot\vec r_{ab}\times\vec p_a\bigr)+
\bigl(\vec\sigma_b\cdot\vec r_{ab}\times\vec p_b -
\vec\sigma_a\cdot\vec
r_{ab}\times\vec p_a\bigr)\biggr]\biggr\}\,.\label{10}
\end{eqnarray}
For states with vanishing angular momentum $L$ and spin $S=1/2$, 
the expectation value is simplified to the form
\begin{eqnarray}
E^{(4)} &=& \langle\Psi|H^{(4)}|\Psi\rangle =
\biggl\langle\sum_a \biggl\{-\frac{\vec p^{\,4}_a}{8\,m^3} +
\frac{ \pi\,Z\,\alpha}{2\,m^2}\,\delta^3(r_a)\biggr\}
\nonumber \\
&& +\sum_{a>b}\sum_b \biggl\{
\frac{ \pi\,\alpha}{m^2}\, \delta^3(r_{ab})
-\frac{\alpha}{2\,m^2}\, p_a^i\,
\biggl(\frac{\delta^{ij}}{r_{ab}}+\frac{r^i_{ab}\,r^j_{ab}}{r^3_{ab}}
\biggr)\, p_b^j \biggr\}\biggr\rangle\,.
\label{11}
\end{eqnarray} 
$E^{(4)}$ has already been obtained in works \cite{yan_delta, king_delta}.
Calculations of these matrix elements involves the usual Hylleraas integrals
with all $n_i$ nonnegative and extended integrals, namely with one 
parameter $n_i$ equal to $-1$. The direct numerical method to calculate these
integrals was presented in \cite{king_lit, yan_lit1}. Here we apply the analytic approach.
Recursion relations for the case
of $n_1$ or $n_2$ or $n_3$ equal to $-1$ have been obtained in \cite{rec_sing}.
Hylleraas integrals involving $n_4$ or $n_5$ or $n_6$ equal to $-1$
can in principle be obtained by the integration of 
the usual Hylleraas integral with respect to the corresponding parameter
$w_i$ \cite{rec_sing}. However, some recursion relations may become unstable,
for example in the case of $n_4=-1$ the recursion in $n_1$ is numerically 
unstable for large $w_1$. To avoid this problem we derive in the next section
stable recursion relations for extended Hylleraas integrals with $n_i=-1$ for $i=4,5,6$.
Numerical results for matrix elements of the Breit Hamiltonian using these
recursion relations, has been presented in Table I and II. 
One observes that the lowest convergence is for the $-p^4/8$ term,
and in spite of the differences for separate matrix elements,
the total relativistic correction is in good agreement with the former 
result in \cite{yan_delta}.

\begin{table}[!hbt]
\caption{Matrix elements of the Breit-Pauli Hamiltonian $H^{(4)}$ in atomic units. }
\label{table2}
\begin{ruledtabular}
\begin{tabular}{rlll}
$\Omega$  &$\sum_a -\frac{1}{8} \nabla_a^4$ &
$\sum_{a>b}\frac{1}{2}\,\nabla^i_a\,\bigl(
\frac{\delta^{ij}}{r_{ab}}+\frac{r^i_{ab}\,r^j_{ab}}{r^3_{ab}}\bigr)\nabla^j_b$
  & $H^{(4)}$ \\ \hline
     $3$  & -78.587\,286\,690\,90 & -0.438\,632\,545\,84 & -12.080\,670\,336\,80 \\
     $4$  & -78.557\,331\,859\,61 & -0.436\,096\,586\,40 & -12.053\,111\,944\,61 \\
     $5$  & -78.556\,355\,905\,97 & -0.435\,697\,344\,91 & -12.050\,709\,116\,55 \\
     $6$  & -78.556\,714\,503\,43 & -0.435\,616\,426\,50 & -12.050\,388\,076\,38 \\
     $7$  & -78.556\,195\,780\,85 & -0.435\,602\,362\,02 & -12.050\,004\,294\,51 \\
     $8$  & -78.556\,162\,642\,13 & -0.435\,599\,523\,90 & -12.049\,961\,162\,16 \\
     $9$  & -78.556\,137\,477\,61 & -0.435\,598\,217\,44 & -12.049\,926\,149\,76 \\
     $10$ & -78.556\,135\,734\,01 & -0.435\,598\,047\,58 & -12.049\,921\,414\,27 \\
     $11$ & -78.556\,131\,596\,34 & -0.435\,597\,963\,57 & -12.049\,916\,800\,81 \\
     $12$ & -78.556\,128\,632\,10 & -0.435\,597\,910\,50 & -12.049\,913\,772\,96 \\
     $\infty$ & -78.556\,112\,88(200) & -0.435\,597\,765(50) & -12.049\,897\,86(200)  \\
     Ref. \cite{yan_delta} & -78.556\,135\,55(148) & -0.435\,598\,001\,(137) &  -12.049\,909\,94(180) \\
\end{tabular}
\end{ruledtabular}
\end{table}

\section{Recursion relations for three-electron extended Hylleraas integral
         with $1/r_1^2$}
In the former section we calculated relativistic corrections.
For this we needed various extended Hylleraas integrals,
among them, integrals with $1/r_i^2$, which are being derived here.
To obtain recursion relations for three-electron Hylleraas
integral in Eq. (\ref{07}), one first considers the integral $G$
\begin{eqnarray}
G(m_1,m_2,m_3;m_4,m_5,m_6) &=& \frac{1}{8\,\pi^6}\,\int d^3k_1\int d^3k_2\int d^3k_3\,
(k_1^2+u_1^2)^{-m_1}\,(k_2^2+u_2^2)^{-m_2} \nonumber\\ &&\hspace*{-1cm}(k_3^2+u_3^2)^{-m_3}\,
(k_{32}^2+w_1^2)^{-m_4}\,(k_{13}^2+w_2^2)^{-m_5}\,(k_{21}^2+w_3^2)^{-m_6}, \label{12}
\end{eqnarray}
which is related to $f$ by: $f(0,0,0,0,0,0) = G(1,1,1,1,1,1)|_{u_1=u_2=u_3=0}$.
The following 9 integration by part identities are valid
because the integral of the derivative of a function 
vanishing at infinity vanishes,
\begin{eqnarray}
&&0 \equiv {\rm id}(i,j) = 
\int d^3k_1\int d^3k_2\int d^3k_3\,\frac{\partial}{\partial\,{\vec k_i}}
 \Bigl[ \vec k_j\,(k_1^2+u_1^2)^{-1} 
\nonumber \\ &&  
(k_2^2+u_2^2)^{-1}\,(k_3^2+u_3^2)^{-1}
(k_{32}^2+w_1^2)^{-1}\,(k_{13}^2+w_2^2)^{-1}\,(k_{21}^2+w_3^2)^{-1} 
\Bigr] ,
\label{13}
\end{eqnarray} 
where $i,j=1,2,3$.
The reduction of the scalar products from the numerator leads to the 
identities for the linear combination of the $G$ functions.
If any of the arguments is equal to 0, then $G$ becomes a known two-electron
Hylleraas type integral. These identities are used to derive various
recursion relations. Here, we derive a set of recursions for the case when $n_4$, $n_5$ 
or $n_6$ is equal to $-1$. Let us assume that $n_4=-1$. The analytic expression 
for $f(0,0,0,-1,n_5,n_6)$ involves powers of $w_2-w_3$ in the denominator which is
not very convenient in high precision numerical calculations.
Instead, we use recursions for  $f(0,0,0,0,n_5,n_6)$ and numerically integrate
with respect to $w_1$, namely
\begin{equation}
f(0,0,0,-1,n_5,n_6) = \int_{w_1}^\infty d\,w_1\,f(0,0,0,0,n_5,n_6)\,.
\label{14}
\end{equation}
These recursions are derived as follows. We take ${\rm id}(i,i)$ with $i=1,2,3$
and put $u_i=0$. Resulting three equations are solved against three unknowns:
$G(1,1,1,2,1,1)$, $G(1,1,1,1,2,1)$, and $G(1,1,1,1,1,2)$.
The solution for the last two $G$ functions is the following
\begin{eqnarray}
G(1,1,1,1,2,1) &=&  \frac{1}{w_2^2}\,\bigl[G(0, 1, 1, 1, 1, 2) - G(1, 0, 1, 1, 1,2)  - G(1, 0, 1, 2, 1, 1) 
\nonumber \\ && 
+ G(1, 1, 0, 2, 1, 1) + G(1, 1, 1, 1, 1, 1)/2\bigr]\,,\label{15}\\ 
G(1,1,1,1,1,2) &=&  \frac{1}{w_3^2}\,\bigl[G(0, 1, 1, 1, 2, 1) + G(1, 0, 1, 2, 1, 1) - G(1, 1, 0, 1, 2, 1) 
\nonumber \\ && 
- G(1, 1, 0, 2, 1, 1) + G(1, 1, 1, 1, 1, 1)/2 \bigr]\,.\label{16}
\end{eqnarray}
By differentiation with respect to $w_2$ and $w_3$ one obtains the following
recursion relations
\begin{eqnarray}
f(0, 0, 0, 0, n_5+1, n_6) &=& \frac{1}{w_1\,w_2\,w_3}\,
 \bigl[ (n_5+1)\,f(0, 0, 0, 0, n_5, n_6)\,w_1\,w_3
\nonumber \\ &&
- (n_5+1)\,n_6\,f(0, 0, 0, 0, n_5, n_6-1)\,w_1 
\nonumber \\ && 
+n_6\,f(0, 0, 0, 0, n_5+1, n_6-1)\,w_1\,w_2 
\nonumber \\ && 
-n_6\,\Gamma(n_5, n_6-1, -1, w_1 + w_2, w_3, 0) 
\nonumber \\ && 
+n_6\,\Gamma(n_6-1, n_5, -1, w_1 + w_3, w_2, 0) 
\nonumber \\ && 
-\Gamma(n_6, n_5, -1, w_1 + w_3, w_2, 0)\,w_1 
\nonumber \\ && 
+\Gamma(n_5 + n_6, 0, -1, w_2 + w_3, w_1, 0)\,w_1 
\nonumber \\ && 
+\Gamma(n_5, n_6, -1, w_1 + w_2, w_3, 0)\,w_3 
\nonumber \\ && 
-\Gamma(n_6, n_5, -1, w_1 + w_3, w_2, 0)\,w_3\bigr]\,,
\label{17}
\end{eqnarray}
\begin{eqnarray}
f(0, 0, 0, 0, n_5, n_6+1) &=& \frac{1}{w_1\,w_2\,w_3}
 \bigl[(n_6+1)\,f(0, 0, 0, 0, n_5, n_6)\,w_1\,w_2 
\nonumber \\ &&
- n_5\,(n_6+1)\,f(0, 0, 0, 0, n_5-1, n_6)\,w_1 
\nonumber \\ &&
+n_5\,f(0, 0, 0, 0, n_5-1, n_6+1)\,w_1\,w_3
\nonumber \\ &&
+n_5\,\Gamma(n_5-1, n_6, -1, w_1 + w_2, w_3, 0) 
\nonumber \\ &&
-n_5\,\Gamma(n_6, n_5-1, -1, w_1 + w_3, w_2, 0) 
\nonumber \\ &&
-\Gamma(n_5, n_6, -1, w_1 + w_2, w_3, 0)\,w_1 
\nonumber \\ &&
+\Gamma(n_5 + n_6, 0, -1, w_2 + w_3, w_1, 0)\,w_1 
\nonumber \\ &&
-\Gamma(n_5, n_6, -1, w_1 + w_2, w_3, 0)\,w_2 
\nonumber \\ &&
+\Gamma(n_6, n_5, -1, w_1 + w_3, w_2, 0)\,w_2 \bigr]\,.
\label{18}
\end{eqnarray}
where $\Gamma$ is a known \cite{gamma1, gamma2,gamma3} two-electron integral
\begin{equation}
\Gamma(n_1,n_2,n_3,\alpha_1,\alpha_2,\alpha_3) = 
\int\frac{d^3\,r_1}{4\,\pi}
\int\frac{d^3\,r_2}{4\,\pi}\,e^{-\alpha_1\,r_1-\alpha_2\,r_2-\alpha_3\,r_{12}}\,
r_{1}^{n_1-1}\,r_{2}^{n_2-1}\,r_{12}^{n_3-1}\,. 
\end{equation}
The integration in Eq. (\ref{14}) is performed numerically
using adapted points and weights to the function
which has logarithmic end-point singularity,
namely
\begin{equation}
\int_0^1 dx\,\bigl[W_1(x) + W_2(x)\,\ln(x)\bigr]\,,
\label{19}
\end{equation}
where $W_i$ are functions without any singularities.
The method to obtain $n$ adapted points and weights is presented in Appendix A,
and this integral is exact for $W_i$ being polynomials up to the  order $n-1$.
In the actual calculations we achieved at least 48 digits precision
with only 100 points. Having obtained $f(0,0,0,-1,n_5,n_6)$
we construct recursion relations in $n_1$, $n_2$, and $n_3$.
This is achieved in two steps. In the first step we use
integration by parts in momentum representation Eq. (\ref{13}), to form the following
linear combination
\begin{eqnarray}
{\rm id}(2,2)+{\rm id}(3,3)-{\rm id}(1,1) &=& 2\,\bigl[
G(0, 1, 1, 1, 1, 2) + G(0, 1, 1, 1, 2, 1) - G(1, 0, 1, 1, 1, 2) 
\nonumber \\ &&
- G(1, 1, 0, 1, 2, 1) - G(1, 1, 1, 1, 1, 1)/2 - G(2, 1, 1, 1, 1, 1)\,u_1^2 
\nonumber \\ &&
-G(1, 1, 1, 1, 1, 2)\,(u_1^2 - u_2^2) + G(1, 2, 1, 1, 1, 1)\,u_2^2 
\nonumber \\ &&
- G(1, 1, 1, 1, 2, 1)\,(u_1^2 - u_3^2) + G(1, 1, 2, 1, 1, 1)\,u_3^2 
\nonumber \\ &&
+ G(1, 1, 1, 2, 1, 1)\,w_1^2\bigr] = 0\,.
\label{20}
\end{eqnarray} 
We integrate with respect to $w_1$ and differentiate over $u_1$, $u_2$, $u_3$,
$w_2$, and $w_3$ to obtain the main formula
\begin{eqnarray}
f(n_1, n_2, n_3, -1, n_5, n_6) &=& \frac{1}{(n_2 + n_3 - n_1)\,w_2\,w_3}\,\bigl[ \nonumber \\ && 
(n_1 - 1)\,n_1\,n_5\,f(n_1-2, n_2, n_3, -1, n_5-1, n_6+1) \nonumber \\ && 
+ (n_1 - 1)\,n_1\,n_6\,f(n_1-2, n_2, n_3, -1, n_5+1, n_6-1) \nonumber \\ && 
-(n_2 - 1)\,n_2\,n_5\,f(n_1, n_2-2, n_3, -1, n_5-1, n_6+1) \nonumber \\ && 
- (n_3 - 1)\,n_3\,n_6\,f(n_1, n_2, n_3-2, -1, n_5+1, n_6-1) \nonumber \\ && 
+(n_1 - n_2 - n_3)\,n_5\,n_6\,f(n_1, n_2, n_3, -1, n_5-1, n_6-1) \nonumber \\ && 
+ n_5\,n_6\,f(n_1, n_2, n_3, 0, n_5-1, n_6-1)\,w_1 \nonumber \\ && 
- (n_1 - 1)\,n_1\,f(n_1-2, n_2, n_3, -1, n_5, n_6+1)\,w_2 \nonumber \\ && 
+ (n_2 - 1)\,n_2\,f(n_1, n_2-2, n_3, -1, n_5, n_6+1)\,w_2 \nonumber \\ && 
- (n_1 - n_2 - n_3)\,n_6\,f(n_1, n_2, n_3, -1, n_5, n_6-1)\,w_2 \nonumber \\ && 
- n_6\,f(n_1, n_2, n_3, 0, n_5, n_6-1)\,w_1\,w_2 \nonumber \\ && 
- (n_1 - 1)\,n_1\,f(n_1-2, n_2, n_3, -1, n_5+1, n_6)\,w_3 \nonumber \\ && 
+ (n_3 - 1)\,n_3\,f(n_1, n_2, n_3-2, -1, n_5+1, n_6)\,w_3 \nonumber \\ && 
- (n_1 - n_2 - n_3)\,n_5\,f(n_1, n_2, n_3, -1, n_5-1, n_6)\,w_3 \nonumber \\ && 
- n_5\,f(n_1, n_2, n_3, 0, n_5-1, n_6)\,w_1\,w_3 \nonumber \\ && 
+ f(n_1, n_2, n_3, 0, n_5, n_6)\,w_1\,w_2\,w_3 \nonumber \\ && 
+ n_6\,\delta(n_3)\,\Gamma(n_5-1, n_6-1, n_1 + n_2 - 1, w_1 + w_2, w_3, 0)\nonumber \\ &&  
+ n_5\,\delta(n_2)\,\Gamma(n_6-1, n_5-1, n_1 + n_3 - 1, w_1 + w_3, w_2, 0) \nonumber \\ && 
- n_5\,\delta(n_1)\,\Gamma(n_5 + n_6-1, -1, n_2 + n_3 - 1, w_2 + w_3, w_1, 0) \nonumber \\ && 
- n_6\,\delta(n_1)\,\Gamma(n_5 + n_6-1, -1, n_2 + n_3 - 1, w_2 + w_3, w_1, 0) \nonumber \\ && 
- \delta(n_2)\,\Gamma(n_6, n_5-1, n_1 + n_3 - 1, w_1 + w_3, w_2, 0)\,w_2 \nonumber \\ && 
+ \delta(n_1)\,\Gamma(n_5 + n_6, -1,  n_2 + n_3 - 1, w_2 + w_3, w_1, 0)\,w_2 \nonumber \\ && 
- \delta(n_3)\,\Gamma(n_5, n_6-1, n_1 + n_2 - 1, w_1 + w_2, w_3, 0)\,w_3 \nonumber \\ && 
+ \delta(n_1)\,\Gamma(n_5 + n_6, -1, n_2 + n_3 - 1, w_2 + w_3, w_1, 0)\,w_3 \bigr]\,.
\label{21}
\end{eqnarray}
This general formula does not work in the case $n_1=n_2+n_3$.
In the second step we use integration by part identities in the coordinate
space to fill this hole. We limit ourselves only to a special case
of these identities in the form
\begin{equation}
0 = {\rm id}(i) \equiv \int d^3 r_1\,\int d^3 r_2\,\int d^3 r_3\,
\bigl(g\,\nabla^2_i h - h\,\nabla^2_i g\bigr)\,,
\label{22}
\end{equation}
where
\begin{eqnarray}
g &=& e^{-w_1\,r_1-w_2\,r_2-w_3\,r_3}\,r_1^{n_4-1}\,r_2^{n_5-1}\,r_3^{n_6-1}\,,
\nonumber \\
h &=& r_{23}^{n_1-1}\,r_{31}^{n_2-1}\,r_{12}^{n_3-1}\,.
\label{23}
\end{eqnarray}
The identities id$(2)$ and id$(3)$ 
\begin{eqnarray}
f(n_1, n_2, n_3, -1, n_5, n_6) &=& 
\bigl[(n_1-1)\,(n_1 + n_3-1)\,f(n_1-2, n_2, n_3, -1, n_5, n_6) \nonumber \\ &&
-(n_1-1)\,(n_3-1)\,f(n_1-2,n_2+2,n_3-2, -1, n_5, n_6)  \nonumber \\ &&
+ (n_3-1)\,(n_1 + n_3-1)\,f(n_1, n_2, n_3-2, -1, n_5, n_6)  \nonumber \\ &&
-(n_5-1)\,n_5\,f(n_1, n_2, n_3, -1, n_5-2, n_6)  \nonumber \\ &&
+2\,n_5\,f(n_1, n_2, n_3, -1, n_5-1, n_6)\,w_2   \nonumber \\ &&
+ \delta(n_5)\,\Gamma(n_1 + n_6-1, n_3-2, n_2, w_3, w_1, 0)\bigr]/w_2^2\,,  \label{24}\\ 
f(n_1, n_2, n_3, -1, n_5, n_6) &=&
\bigl[-(n_1-1)\,(n_2-1)\,f(n_1-2, n_2-2, n_3+2, -1, n_5, n_6)  \nonumber \\ &&
+ (n_1-1)\,(n_1 + n_2-1)\,f(n_1-2, n_2, n_3, -1, n_5, n_6) \nonumber \\ && 
+(n_2-1)\,(n_1 + n_2-1)\,f(n_1, n_2-2, n_3, -1, n_5, n_6)  \nonumber \\ &&
- (n_6-1)\,n_6\,f(n_1, n_2, n_3, -1, n_5, n_6-2)  \nonumber \\ &&
+ 2\,n_6\,f(n_1, n_2, n_3, -1, n_5, n_6-1)\,w_3   \nonumber \\ &&
+ \delta(n_6)\,\Gamma(n_2 -2, n_1 + n_5-1, n_3, w_1, w_2, 0)\bigr]/w_3^2\,, \label{25}
\end{eqnarray}
replace the main recursion in Eq. (\ref{21}) for the case $n_1=n_2+n_3$ and can be
used also for all other $n_i$ under conditions that $n_1>0$, $n_3>0$ or
$n_1>0$, $n_2>0$, respectively.

\section{Summary}
We have demonstrated the advantages of the analytic approach to three-electron
Hylleraas integrals by the calculation of nonrelativistic energy of the ground
state lithium atom and the leading relativistic corrections. The achieved
accuracy is the best to date and this is mainly due to
the use of much larger basis sets. In fact it is possible to perform calculation
with $\Omega>12$ by using sextuple  precision arithmetics.
The typical evaluation time in sextuple precision for $\Omega=12$ 
is 24 hours on 2.4 GHz Opteron, and most of the time is devoted to 
LU decomposition.  

Having precise wave functions, we have calculated leading relativistic
corrections and the results only partially agree with that of Yan and Drake 
\cite{yan_delta} and of King \cite{king_delta}.
We are now in position to calculate higher order, namely $m\,\alpha^6$
relativistic and QED corrections, for example to the lithium
ground state hyperfine splitting \cite{lit_hfs}. However, this involves
more complicated Hylleraas integrals containing two factors
among $1/r_i^2$ and $1/r_{ij}^2$, which have not yet been worked out by
the recursion methods of the authors.

Even more interesting is the possible extension of this analytic method to
beryllium and beryllium-like ions, the 4-electron systems. 
The use of large Hylleraas basis set will allow for 
a high precision calculation of the wave function, energies and
transition rates. For example, knowing the isotope shifts, one can 
obtain charge radii as for the lithium isotope \cite{lit_iso}. 
General Hylleraas integrals
for 4-electron systems has not yet been worked out \cite{beryl2, beryl4}. The so called
double linked basis set, the functions with at most two odd powers of
$r_{ij}$ have been used by B\"usse {\em et al} in \cite{beryl3} to obtain an accurate
nonrelativistic energy, but still less accurate than the result of Komasa in
\cite{komasa_be}. It has not yet been attempted to calculate relativistic corrections
with Hylleraas functions as they involve even more difficult integrals.
We think, the integration by part technique, should allow
for the derivation of compact formulas for all 4-electron 
Hylleraas integrals.

Our primary motivation for developing Hylleraas basis set is 
the calculation of higher order relativistic and QED effects, and
to demonstrate that standard techniques used in relativistic quantum
chemistry, which are based on the multi-electron Dirac-Coulomb Hamiltonian  are not correct
for principal reasons. This Hamiltonian does not not include properly negative energy states.
The correct treatment has to be based on quantum electrodynamics
and several very accurate results for few electron ions
have already been obtained within the so called $1/Z$ expansion \cite{heavy1,
  heavy2, heavy3}. Nevertheless, there is no yet formalism which allows for 
systematic inclusion of  negative energy states and QED effects for
many electron atoms.

\section{Acknowledgments}
We are grateful to Vladimir Korobov for his source code
of the fast multi-precision arithmetics and to Micha\l\ Bernardelli
for bringing the work \cite{rokhlin} to our attention.
This work was supported by EU Grant No. HPRI-CT-2001-50034.

\appendix
\renewcommand{\theequation}{\Alph{section}\arabic{equation}}
\renewcommand{\thesection}{\Alph{section}}

\section{Quadrature with logarithmic end-point singularity}
Consider the integral
\begin{equation}
I = \int_0^1 dx\,\bigl[W_1(x) + \ln(x)\,W_2(x)\bigr]\,,
\label{A1}
\end{equation}
where $W_i$ are arbitrary polynomials of maximal degree $n-1$.
We would like to find $n$ nodes $x_i$ and $n$ weights $w_i$ such that
\begin{equation} 
I = \sum_{i=1}^n\,w_i\,\bigl[W_1(x_i) + \ln(x_i)\,W_2(x_i)\bigr]\,.
\label{A2}
\end{equation}
In general it is a difficult numerical problem to find
a solution of corresponding $2\,n$ nonlinear equations with $j=1,n$
\begin{eqnarray}
\int_0^1 dx\,x^{j-1} = \frac{1}{j} &=&  \sum_{i=1}^n\,w_i\,
x_i^{\,j}\,, \label{A3}\\
\int_0^1 dx\,x^{j-1}\,\ln x = -\frac{1}{j^2} &=&  \sum_{i=1}^n\,w_i\,
x_i^{\,j}\,\ln x_j\,. \label{A4}
\end{eqnarray}
The work \cite{rokhlin} solves this problem and proves that $w_i$ are all positive.
The solution is as follows. 

One defines $2\,n$ functions $\phi_i$
\begin{eqnarray}
\phi_k(x) &=& x^{k-1},\;{\rm for}\; k=1,n \label{A5}\\
\phi_k(x) &=& x^{k-1}\,\ln x,\;{\rm for}\; k=n+1,2\,n\,. \label{A6}
\end{eqnarray}
Consider $n$ points $x_i$ which are not necessarily the solution of
Equations (\ref{A3},\ref{A4}) but are close to them, and construct another set of functions
$\sigma_i$, $\eta_i$, for $i=1,n$
\begin{eqnarray}
\sigma_i(x) = \sum_{j=1}^{2\,n}\alpha_{ij}\,\phi_j(x)\,,\label{A7}\\
\eta_i(x) = \sum_{j=1}^{2\,n}\beta_{ij}\,\phi_j(x)\,, \label{A8}
\end{eqnarray}
such that
\begin{eqnarray}
\sigma_i(x_k) &=& 0\,,\nonumber \\
\sigma'_i(x_k) &=& \delta_{ik}\,, \nonumber \\
\eta_i(x_k) &=& \delta_{ik}\,, \nonumber \\
\eta'_i(x_k) &=& 0\,. \label{A9}
\end{eqnarray}
The set of conditions (\ref{A9}) uniquely determines the matrices $\alpha_{ij}$ and
$\beta_{ij}$. If $x_k$ are nodes, then
\begin{eqnarray}
\int_0^1 dx\,\sigma_i(x) &=& 0\,,\nonumber \\
\int_0^1 dx\,\eta_i(x) &=& w_i\,. \label{A10}
\end{eqnarray} 
If $x_k$ are not exactly the nodes, but are sufficiently close, then
according to work \cite{rokhlin}, the iteration $x_i\rightarrow \tilde x_i$
\begin{equation}
\tilde x_i = x_i + \frac{\int_0^1 dx\,\sigma_i(x)}{\int_0^1 dx\,\eta_i(x)}\,,
\label{A11}
\end{equation}
converges to nodes, the solution of Eqs. (\ref{A3},\ref{A4}). The only problem now, is to find a
sufficiently good initial values for $x_i$. For this one constructs
a homotopy $\phi_k(x,t)$ such that
\begin{eqnarray}
\phi_k(x,t) &=& x^{k-1}\;{\rm for}\; k=1,n \,,\nonumber\\
\phi_k(x,t) &=& (1-t)\,\sqrt{x} + t\,x^{k-1-n}\,\ln (x) \;{\rm for}\; k=n+1,2\,n\,.
\label{A12}
\end{eqnarray}
At $t=0$, $\phi(x,0)$ are polynomials in $\sqrt{x}$, therefore
one obtains $x_i = y_i^2$ where $y_i$ are nodes for Gauss-Legendre quadrature.
By slowly changing $t$ from $0$ one finds the solution at $t=1$.
In the actual numerical calculations we found that the steps
$t_i = i/100$ were sufficiently small for the above iteration to converge. 
This generalized Gaussian quadrature can also be constructed for other types
of functions including various, even nonintegrable singularities. 

\end{document}